\documentclass{aa}  

\usepackage{graphicx}
\usepackage{txfonts}
\usepackage{xcolor}
\usepackage[normalem]{ulem}

\newcommand{\kms}{km s$^{-1}$}
\newcommand{\VphiLim}{170}

\newcommand{\kmskpc}{km s$^{-1}$ kpc$^{-1}$}
\newcommand{\FeH}{[\mathrm{Fe/H}]}

\newcommand{\rrc} {\mbox{RR\emph{c}}}

\begin{document}

   \title{RR Lyrae stars trace the Milky Way warp}


   \author{Mauro Cabrera-Gadea
          \inst{1}
          \and
          Cecilia Mateu\inst{1} 
          \and 
          Pau Ramos \inst{2}
          }

   \institute{$^1$ Departamento de Astronom\'{i}a, Instituto de F\'{i}sica, Universidad de la Rep\'{u}blica, Igu\'a 4225, CP 11400 Montevideo, Uruguay\\
   $^2$ National Astronomical Observatory of Japan, Mitaka-shi, Tokyo 181-8588, Japan\\
              \email{mauro.cabrera@pedeciba.edu.uy}
         }

   \date{Received ...; accepted ...}

 
  \abstract
   {The outskirts of the Milky Way disc have been known to be warped since the late 1950s. Although various stellar populations have shown an underlying warped distribution, the relation between the age of the population and the warp they trace remains an open question. Understanding this relation may shed light on the origin of the warp which remains a puzzle to be solved.}
   {Our goal in this work is to detect the presence of the warp in the RR Lyrae (RRL) population of the Milky Way disc.} 
   {We use a compilation of the three largest public catalogues of RRL stars, precise photometric distances ($\sim5\%$) and Gaia DR3 proper motions to kinematically select a sample of thin disc RRL in the Galactic anticentre, where the tangential velocity best approximates the azimuthal velocity to differentiate between those that rotate (disc) and those that do not (halo). For those disc-like RRL we analyse their mean vertical height and mean vertical velocity.
   }
   {We show, for the first time, that RRL stars with thin disc-like kinematics trace the warp. 
   In the anticentre direction, the RRL population reaches a minimum in mean vertical height of $\approx0.4$~kpc, with a trend systematically lower than the one found with Classical Cepheids. The kinematical signal of the RRL warp starts at $R\approx10$~kpc and, rather than resembling the Cepheid's, shows a similar trend to the Red Clump population from previous works, reaching a maximum value of $\approx7$~\kms in vertical velocity. We also obtain an estimation of the pattern speed of the RRL warp with a prograde rotation of $\approx13\pm2$~\kmskpc, compatible with results obtained from Classical Cepheids. Finally, we also obtain a vertical velocity dispersion $\approx 17$~\kms, inconsistent with the kinematics of a canonical old age ($>10$~Gyr) disc population and, instead, favouring a population dominated by intermediate-age (3-4 Gyr).}
   {Our results indicate that the thin disc RRL stars are a dynamical intermediate-age tracer of the warp, opening a new window to study the dependency of the warp with stellar age in the Milky Way.}


   \maketitle
%

\section{Introduction}

It has been known for a long time that the Milky Way disc is warped \citep{Burke_HI_Warp_1957}. Different disc tracers have shown features in their positions and/or kinematics typical of a warped disc \citep{Warp_Pulsars_Yusifov_2004,Levine2006,Skowron2019,Chen3Dmap,RG19,Poggio_Warp_evolving,Cheng_2020_Warp_patern,Li_2023_Nonsteady_Warp,Dehnen_2023_Warp_Ceph,McMillan_Wap_2024}. Although stellar populations with different ages have been found to trace a warped disc, the relationship between the warp's shape and the age of the tracer is still unclear. Some works have found the amplitude of the warp increases for older stellar populations \citep{Evolution_disc_shape_Amores_2017,RG19} while others claim the opposite \citep{Cheng_2020_Warp_patern,Chrobacova_GDR2_Warp_Model_2020GDR2,Wang_LAMOST_Disk_Warp_RC_2020,Li_2023_Nonsteady_Warp}. The differences and contradictions in the literature may arise from several factors like the purity of the samples, their selection function and also the uncertainties in the estimation of distances in the outer regions of the disc where the warp is present. The extinction in the disc affects mainly the detection of stars close to the Galactic plane, which can lead to an overestimation of the amplitude of the warp \citep[see Appendix B in][]{Cabrera_Gadea_etal_2024}; purity affects the reliability of the age determination, and accuracy and precision in the distance determination are key to study the global warp structure and kinematics. 

Standard candles offer the best performance in the determination of distance and, generally, ensure the purity of the sample. Classical Cepheids have been widely used to study the warp because of their excellent distance precision ($\sim$3\%), wide disc coverage and, since they are all younger than 500~Myr, they act as tracers of the gaseous disc in which they formed. Thus, Cepheids have proved to be of great use in characterising the warp's structure and kinematics \citep{Chen3Dmap,Skowron2019,Dehnen_2023_Warp_Ceph,Cabrera_Gadea_etal_2024}, but they do not help us understand the changes with stellar age. To do that, we would need an alternative standard candle representative of the dynamics of the older stellar component of the thin disc (TnD). To the best of our knowledge, the only previous study looking for the warp with standard candles older than the Cepheids is \cite{Mira_OGLE_2023_Iwanek}, where Mira stars were used. However, as pointed out by the authors, the presence or absence of the warp could not be determined with their sample.

RR Lyrae stars (RRL), are a popular standard candle for the eldest stellar populations (>10~Gyr) and have been widely used as tracers of the thick disc \citep{Layden1995a,Kinemuchi2006,Mateu_thick_disc_2018}. Given its short radial scale length \citep[$h_R=2.1$~kpc,][]{Mateu_thick_disc_2018}, it is not expected that its outermost region would trace the warp since only $\sim2\%$ of thick disc stars would be found at distances beyond the warp's onset ($R\approx10-12$~kpc). However, evidence for RRLs with TnD kinematics has been accumulating \citep{Prudil2020,Zinn2020,Marsakov2019}. Recently, \cite{Iorio_RRL_thin_disc_2021} have shown there is a fraction of RRLs with TnD kinematics beyond the Solar neighbourhood up to $R\approx25$~kpc, which raises the question of whether or not this population also traces the warped TnD. Either scenario would be of great interest to restrict the dynamical origin of the warp in the Galaxy.

RRL stars, as the Cepheids, offer low distance uncertainties ($\sim5\%$), large spatial coverage, with the addition of being more numerous and having a well understood selection function \citep{Mateu2024,Mateu2020}. Contrary to the Cepheids, however, they are not unequivocal tracers of the TnD. RRLs trace almost all structures and substructures of the Milky Way, making it impossible to differentiate which structure they belong to just by virtue of being an RRL. This aspect together with the lack of RRL catalogues with extended radial and azimuthal coverage at low latitude have delayed the analysis of the warp, and indeed of the disc, using RRL stars until recently. 

In this paper we use the largest public catalogues of RRL stars to date (Section \ref{s:rrl_sample}) to detect the Galactic disc's warp by selecting TnD samples based on metallicity and on rotation velocity (Section \ref{s:SelecAntiCentre}). In Section~\ref{s:results} we show that the TnD traced by the RRL has the signal of the warp imprinted in the mean $Z$ and $V_z$ as a function of the galactocentric radius, and we compare it to other results in the literature. Our conclusions are summarised in Sec.~\ref{s:conclusion}.


\section{RR Lyrae sample}\label{s:rrl_sample}

In this work, we used a catalogue containing 309,998 stars,  compiled by combining the three largest public RRL surveys: Gaia DR3 Specific Objects Study \citep[SOS,][]{Clementini2023_SOS_GaiaDR3}, ASAS-SN-II \citep{Jayasinghe2019b} and PanSTARRS1 \citep[PS1,][]{Sesar2017c}. We implemented the quality cuts and cross-matching strategy described in Sec.~2 of \citet{Mateu2020} for each catalogue, removed duplicates based on  Gaia's \verb|source_id| and restricted the final catalogue to stars with $\mathrm{RUWE}<1.4$ to ensure the quality of the astrometric parameters, as recommended by \citet{Lindegren_2021_GEDR_Astrometric_Solution}.

Distances to the RRL stars were computed via the $M_G-\mathrm{[Fe/H]}$ relation from \cite{Garofalo2022} using photometric metallicities recalculated by \citet{Li2023} for Gaia~DR3 RRL. For stars without a photometric metallicity estimate, we assigned a metallicity from a normal distribution with mean $-0.65$~dex and standard deviation $0.28$~dex\footnote{As we will show in Sec.~\ref{Sec:Selection} this corresponds to the mean metallicity of kinematic-thin-disc RRL stars.}. This assumption will affect the star's distance and, by extension, its inferred velocity, as we discuss in Sec.~\ref{Sec:Selection}. For the G-band apparent mean magnitude we used \verb|phot_g_mean_mag| and extinctions were calculated using the \citet{Green2019} 3D extinction map. The extinction law coefficient for the Gaia $G$ band was taken from Eq.~A1 of \citet{Ramos2020}, assuming $BP-RP=0.7$ for all RRL stars and an $R_V=3.1$.

\section{Kinematic selection of TnD RRL in the Anticentre}
\label{s:SelecAntiCentre}

The Galactocentric reference frame and cylindrical coordinate system $(R,\phi,z)$ used throughout this work is the same as presented in \citet{Cabrera_Gadea_etal_2024}, with only a slight difference in the solar Galactocentric cartesian velocity, taken here as $(Vx, Vy, Vz) = (11.10, 248.5, 7.25)$~\kms \citep{V_sun_Schonrich_2010,Reid_2020}. In this frame, negative $V_\phi$ corresponds to prograde disc rotation.

We focus our analysis on the anticentre where the azimuthal velocity $V_\phi^*$ calculated solely from the proper motions, i.e. without a line of sight velocity, best approximates the full azimuthal velocity ($V_\phi$). The anticentre sample was selected as all RRL with $165^\circ<\phi<195^\circ$, $9.5 < R/\text{kpc} < 17 $, $|Z|<3$~kpc, $|V_z^*|<70$~\kms and error in $V_z^*$ and $V_\phi^*<40$~\kms. The cuts in $Z$ and $V_z^*$ remove clear outliers and ensure our sample consists only of disc-like RRL. Finally, after visual inspection of all the light curves, we rejected 16 RRLs as misclassified (see Table~\ref{t:rrl_rejected} and details in Appendix~\ref{a:rrl_rejected}).

\subsection{Selecting fast/disc-like rotating RR Lyrae}
\label{Sec:Selection}

\begin{figure*}
	\includegraphics[width=0.95\textwidth]{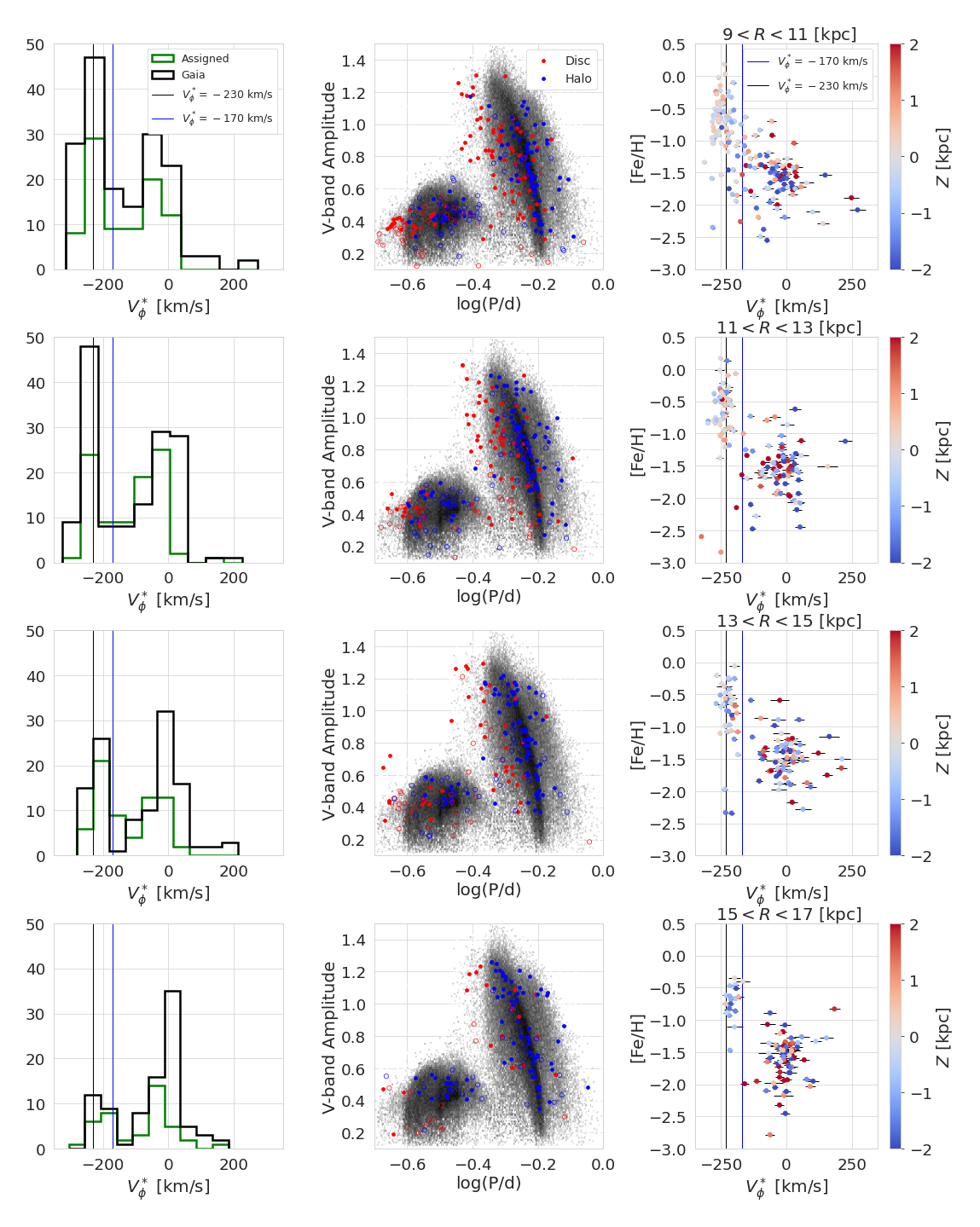}
    \caption{The figure shows the bimodality of the population of RRL stars in velocity and metallicity. The first column shows the histogram of $V_{\phi}^*$ for the RRL with Gaia photometric metallicity (black) and those with metallicity assigned (green, see Sec.~\ref{s:rrl_sample}). The second column shows the period-amplitude (V-band) diagram where red dots are fast rotating RRL with $V_{\phi}^*<-\VphiLim$~\kms and blue dots are slower rotating RRL with $V_{\phi}^*>-\VphiLim$~\kms. 
    The third column shows the $V_\phi^*,\text{[Fe/H]}$ plane, with the black and blue vertical lines indicating $V_\phi^*=-230$~\kms and $V_\phi^*=-\VphiLim$~\kms, respectively for all panels. From top to bottom, each row corresponds to a different radial bin $R\in (9,11),(11,13),(13,15),(15,17)$~kpc.}
    \label{fig:Halo_disc}
\end{figure*}

In Fig.~\ref{fig:Halo_disc} the first column shows histograms for $V_\phi^*$ for the anticentre sample. 
The black and green lines show the histograms for the sample with measured and assigned photometric metallicities (see Sect.~\ref{s:rrl_sample}), respectively. A bimodality is clear in both: a fast component ($V_\phi^*<-\VphiLim$~\kms) rotating like the bulk disc population, containing 321 RRL, and a slower or almost non-rotating component ($V_\phi^*>-\VphiLim$~\kms), corresponding to the halo. Therefore, we define our sample of kinematically-selected TnD RRL stars as those with $V_\phi^*<-\VphiLim$~\kms. The final disc anticentre sample is summarised in Table~\ref{table:disc_rrl_final}.

\begin{table*}
\small
\caption{ Anticentre sample of kinematically-selected thin disc RRL stars}           
\label{table:disc_rrl_final}      
\centering                                    
\begin{tabular}{c c c c c c c c c c c}          
\hline\hline                        
 Source ID & $d$ [kpc]& $\Delta d$ [kpc] & $l$ [deg] & $b$ [deg]& $\mu_l$ [$\mu\text{as/yr}$] & $\mu_b$ [$\mu\text{as/yr}$] & [Fe/H] & P [days] & $\mathrm{Amp}V$ [mag] & F\\    
\hline                                 
     51220203721624064&	3.39& 0.14	&170.018375&	-25.580921&	3.2322&	    -0.5654&    -0.47&	0.481201&	0.84&	0\\      
     72425797289881344&	2.55& 0.13 &152.701939&	-47.293983&	2.7634& 	-0.8368&	-0.35&	0.28662&	0.21&	0\\
     80556926295542528&  1.63& 0.08	& 149.994094 &	-40.849758& 3.3490& 	-8.9908&	-0.63& 0.354279&	0.40&	0\\
\hline                                             
\end{tabular}
\hfill\parbox[t]{\textwidth}{Description of the columns: Gaia DR3 source ID, distance ($d$), distance error ($\Delta d$), galactic longitude ($l$) and latitude ($b$), proper motion in $l$ ($\mu_l$) and $b$ ($\mu_b$), the metallicity adopted for the distance calculation ([Fe/H]), the period (P), V-band amplitude (Amp$V$) and the value "F" is $1$ when the metallicity adopted is the photometric one and $0$ if it was assigned (see Sec.~\ref{s:rrl_sample}) }
\end{table*}

As previously mentioned in Sec.~\ref{s:rrl_sample}, our assumption on disc metallicity for those stars without photometric-metallicity measurements affects the distance and, by extension, the velocity. If a halo star is (wrongly) assigned a disc metallicity, its distance will be underestimated, and its proper motion will translate into a slower velocity. Because we are selecting disc stars as the ones which rotate faster, it is unlikely this assumption will introduce halo contaminants in our disc sample.

Since we do not have photometric metallicities for all the RRLs in our sample, we used the Period-Amplitude diagram to check whether our full kinematic-selected sample is consistent with what is expected for a more metal-rich population compared to the more metal-poor halo. This is shown in the second column of Fig.~\ref{fig:Halo_disc} for different radial bins. The full and empty dots show whether the RRL has photometric or assigned metallicity, respectively. The RRL with $V_\phi^*<-\VphiLim$~\kms (full and empty red dots) have a shorter mean period than those with $V_\phi^*>-\VphiLim$~\kms. This difference is indicative of a higher metallicity in the rotating component and it is also clear in comparison to the full catalogue (grey), which is dominated by the metal-poor halo population.

Finally, for the RRL with photometric metallicities, the third column in Fig.~\ref{fig:Halo_disc} shows the $(V_\phi^*, \text{[Fe/H]})$ plane colour-coded by vertical height. The bimodality in $V_\phi^*$ also corresponds to a bimodality in metallicity, which is clearer in the outermost radial bins ($R>11$~kpc) where the TnD dominates and the thick disc is no longer present due to its shorter scale length. In the $9<R<11$~kpc bin, the transition between the two separate populations (halo and TnD) seems, unsurprisingly, to be bridged by thick disc RRL. As we move to the outer disc it is clear that the fast-rotating and metal-rich population is always close to the Galactic plane ($|Z|<2$)~kpc and the mean tends to go below the Galactic plane (as observed for the warped TnD traced by Cepheids \citealt{Cabrera_Gadea_etal_2024}).

Therefore, the sample selected of RRL in the anticentre direction by $V_{\phi}^*<-170$~\kms is remarkably consistent with what is expected in kinematics, metallicity and height distribution for the Galactic TnD population \citep[e.g.][]{Das_sigmaVz_2024}.

\section{Results and discussion}\label{s:results}

\subsection{The warp traced across the disc}

\begin{figure*}
	\includegraphics[width=\textwidth]{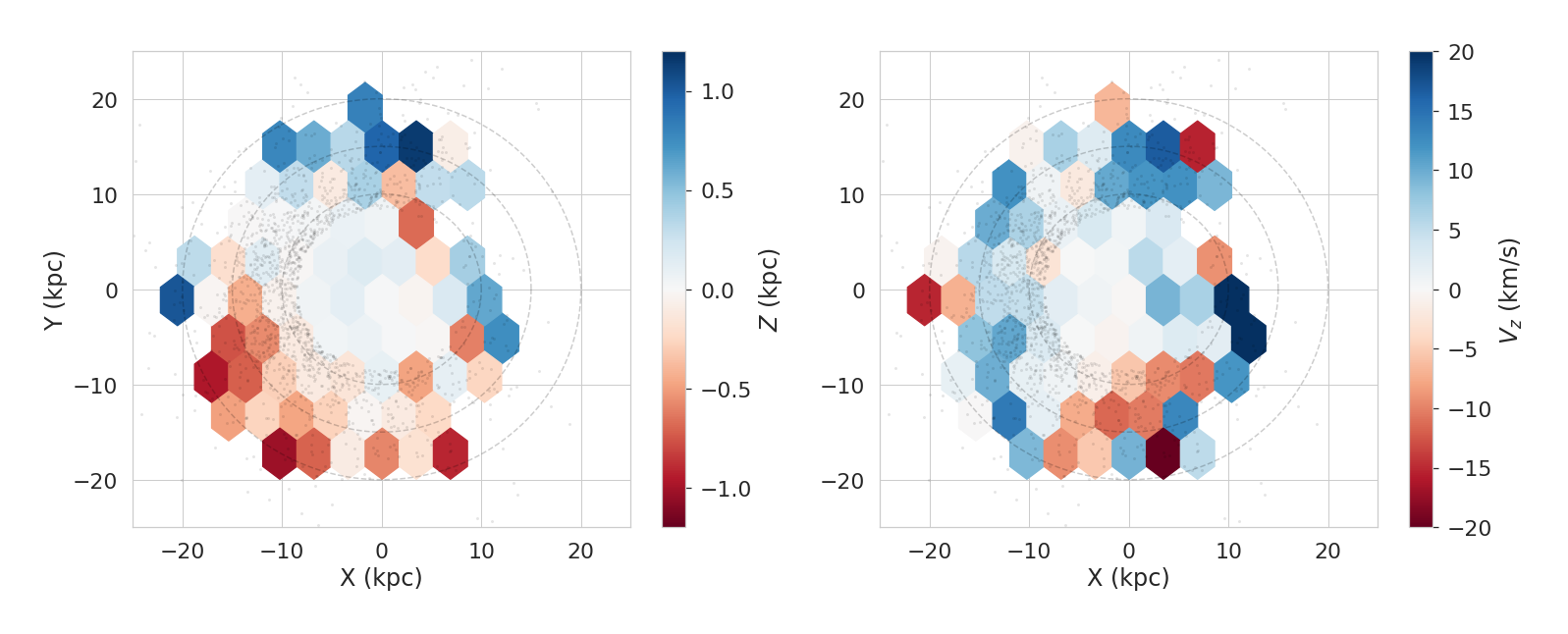}
    \caption{$XY$ plane for the RRL sample with photometric metallicity $\FeH>-1$. In each hexagonal bin the mean $Z$ (left) and $V_z$ (right) computed by a gaussian extreme deconvolution is plotted. The RRL sample with $R>8$~kpc is plotted with black transparent dots. Dashed lines indicate galactocentric rings at $R=10$~kpc, $15$~kpc and $20$~kpc.}
    \label{fig:XY_Z}
\end{figure*}

The last column of plots in Fig.~\ref{fig:Halo_disc} showed that RRLs with $\FeH>-1$ are dominated by the disc population. By selecting stars based on metallicity without filtering by $V_\phi^*$, we can distinguish between the halo and disc populations across all azimuths, without being limited to the anticentre. For this particular sample, Fig.~\ref{fig:XY_Z} presents a map of the mean $Z$ (on the left) and $V_z$ (on the right), obtained by a single component Gaussian extreme deconvolution \citep{astroML,astroMLText} in each bin.

Figure~\ref{fig:XY_Z} (left) shows the underlying warped distribution of the disc RRL population. Since selection effects due to extinction and crowding mostly affect the completeness of RRLs for $X>0$. We have also explored the halo population ([Fe/H]$<-1$) and it does not show any asymmetries (not shown), supporting that the observed trend is likely not an artefact. The sign and coherent pattern, similar to the one traced by classical Cepheids \citep{Skowron2019,Chen3Dmap,Dehnen_2023_Warp_Ceph,Cabrera_Gadea_etal_2024}, clearly reveals the underlying warped distribution.
In the right panel of Fig.~\ref{fig:XY_Z}, the warp's signal appears with positive velocities in the anticentre region as expected from a population ascending towards the northern extreme of the warp. 

Therefore, we have found that the RRL with TnD-like kinematics, and equivalently the most metal-rich RRL, follow a warp across the whole disc, qualitatively similar to the Cepheid's warp. 

\subsection{The warp in the anticentre}

\begin{figure}
	\includegraphics[width=9cm]{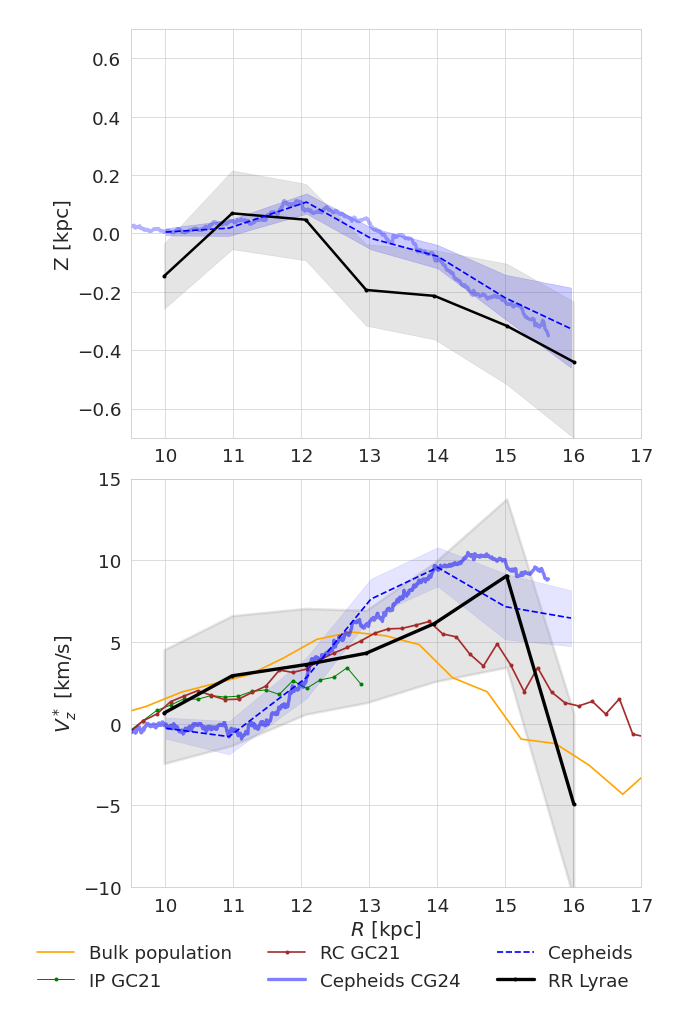}
    \caption{ The vertical height $Z$ as a function of the galactocentric radius $R$ at the anticentre direction for our sample of RRL (black) and Cepheids best-fit warp at the anticentre from \citet[blue solid line]{Cabrera_Gadea_etal_2024}, and for the Cepheids, measured in the same region and with the same method as the RRL (blue dashed line). The shaded areas correspond to one standard deviation from 300 bootstrap realisations. $V_z^*$ as a function of $R$ for RRL stars from this work (black curve), Cepheids results from \citet{Cabrera_Gadea_etal_2024} (blue curve), the measurement for the same region and method as the RRL for the Cepheids sample (blue dashed line), the bulk population from \cite{McMillan_Wap_2024} (yellow curve) and the result from \cite{GEDR3_anticentre_2021} for the red clump (RC, brown curve) and intermediate population (IP, green curve).}
    \label{fig:Vz_Z_lit}
\end{figure}

Using the anticentre sample, we searched for the warp signal up to $R\approx 17$~kpc. 
Figure~\ref{fig:Vz_Z_lit} shows the mean $Z$ (top) and $V_z^*$ (bottom) as a function of radius: in both plots the black lines correspond to the mean in $1$~kpc radial bins, within the 5th and 95th percentiles of $V_z^*$. This result, together with the individual stars, is shown in the Appendix~\ref{apend_AC}.

The top panel in Fig.~\ref{fig:Vz_Z_lit} shows a deviation of mean $Z$ from the galactic midplane by $\approx-0.4$~kpc at $R\approx15$~kpc. Even though $Z$ is more susceptible to bias due to the selection function and extinction effects than $V_z^*$ \citep{RG19,Cabrera_Gadea_etal_2024}, based on the selection function for our RRL catalogue from \citet{Mateu2024}, we are confident the $Z$ trend is robust (see Appendix~\ref{a:RZ_selection_function}). We compare the RRL to our results with Cepheids in \cite{Cabrera_Gadea_etal_2024} (blue curve) and the results for the Cepheids in the same region and method as the RRL (blue dashed curve). We found that the RRL share a similar trend to the Cepheids, going below the Galactic mid-plane, with the RRL being systematically lower than the  Cepheids by $\approx 0.2$~kpc at $R>13$~kpc. Even within the uncertainty (dominated by low number statistics in our sample of RRL), the Cepheid and RRL trends may coincide only marginally at best.

For $V_z^*$ the kinematic signal starts at a shorter radius ($R\approx10$~kpc) than for the Cepheids, for which it begins at $R\approx11$~kpc \citep{Chen3Dmap,Dehnen_2023_Warp_Ceph,Cabrera_Gadea_etal_2024}\footnote{These starting radii may not be the same radius at which the entire warp begins}. The positive trend with the radius reaches a maximum of $\approx 7$~\kms, then declines to $\approx-5$~\kms. As discussed in \citet{Cabrera_Gadea_etal_2024}, this is a clear signal of the warp and similar trends have been found in other stellar populations \citep{Cheng_2020_Warp_patern, Li_2023_Nonsteady_Warp} with different amplitudes but similar in their "arch" shape. The bulk disc population from \cite{McMillan_Wap_2024} (orange) and the result from \cite{GEDR3_anticentre_2021} for red clump stars (RC, brown curve) and intermediate population (IP, green curve)\footnote{To avoid misleading comparisons in our discussion we left out the old population result from \citet{GEDR3_anticentre_2021} because their old population is selected by taking the stars below the turn off of an $8$~Gyr old isochrone. In this selection, low-mass stars of any age coexist and it is not guaranteed that the eldest population dominates the sample.} are also shown.

The difference between the younger and older stellar population in the starting radius and the linear growth in the anticentre direction was observed by \cite{GEDR3_anticentre_2021}. Here it is confirmed by two samples of standard candles with precise distances at different ages. In contrast, we do not observe significant differences between RRL and the intermediate-age populations. For the RC, which can be compared against the RRL sample for a larger range of radii, both populations have a similar mean $V_z^*$ up to $R\approx14$~kpc. For $R>14$~kpc, within our uncertainty, both populations are compatible with the same trend in $V_z^*$. The bulk population from \cite{McMillan_Wap_2024} shows the same trend as the RRL up to $R\approx14$~kpc, and seems to start declining earlier in radius, but at a lower rate.

\subsection{The warp pattern speed}\label{sec:Time_evol}

\begin{figure}
	\includegraphics[width=9cm]{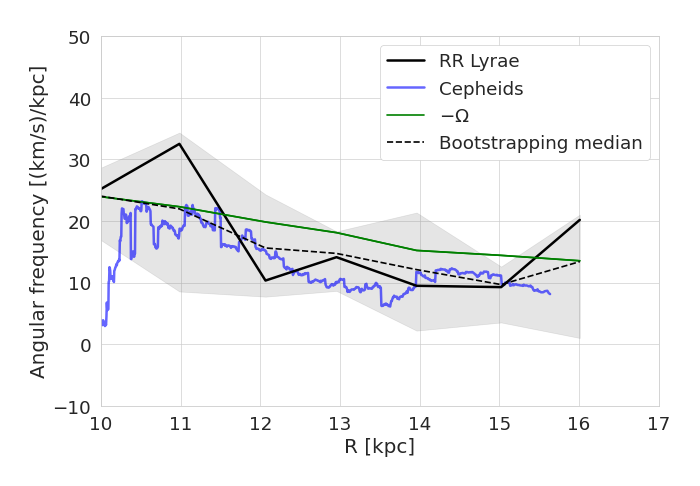}
    \caption{Pattern speed as a function of radius for the RRL's warp (black curve), Cepheid's warp (blue curve) and the angular velocity of the RRL (green curve). The grey shaded area shows the $15\%$ and $85\%$ percentile of the pattern speed results from the bootstrap realisation, the dashed line shows the median of the realisations.}
    \label{fig:omega}
\end{figure}

Inspired by our work in \cite{Cabrera_Gadea_etal_2024}, where we showed that the relative phases of $Z$ and $V_z$ are related to the time evolution of the warp, we found that Fig.~\ref{fig:XY_Z} can give a general picture of the time evolution of the RRL warp. Because of the similarity with Cepheids in the relative phases between $Z$ and $V_z$, i.e. positive $V_z$ where $Z$ goes from negative to positive in the stellar rotation direction, we can expect  the change in amplitude to be negligible ($\dot{A}=0$) and the pattern speed ($\omega$) to be slower than the angular velocity of the stars  \citep[$\Omega$, see Eqs. 14 and 22 in][]{, Cabrera_Gadea_etal_2024}. Under the assumption of $\dot{A}=0$ and a unique pattern speed, we developed a simple method to estimate the instantaneous pattern speed of the warp for our RRL sample (see Apendix~\ref{a:PatternsSpeed}). The results are shown, as a function of the radius, in Fig.~\ref{fig:omega} for the RRL (black curve) and Cepheids  \citep[blue, ][]{Cabrera_Gadea_etal_2024}. The angular velocity ($\langle V_\phi\rangle/R$) of the RRL is shown by the green curve. The RRL warp, thus, shows a prograde rotation with a mean pattern speed of $\approx13\pm2$~\kmskpc for $R>12$~kpc.

\

The pattern speed estimated for the RRL is similar in magnitude and direction to that of the Cepheids \citep{Dehnen_2023_Warp_Ceph,Cabrera_Gadea_etal_2024}, which is also similar to the results by \citet{Poggio_Warp_evolving,Cheng_2020_Warp_patern,He_warp_Ocluster_w_2023,McMillan_Wap_2024,Poggio_great_wave_2024} with other disc tracers. It also shows the differential rotation present in the Cepheids \citep{Dehnen_2023_Warp_Ceph, Cabrera_Gadea_etal_2024, Poggio_great_wave_2024} and young giants \citep{Poggio_great_wave_2024}, but due to our low number statistics, this trend should be taken with caution.

\subsection{The age of the thin disc RR Lyrae}

In order to use RRL for the study of the warp's age dependency, we should assess whether our disc RRL sample is actually representative of an old population ($>10$~Gyr). Although RRL are always assumed to be an old stellar population, for the TnD
they have been proposed to be younger ($5-7$~Gyr) by \citet{Iorio_RRL_thin_disc_2021}\footnote{Similar results were found from completely independent arguments and data by \citet{Sarbadhicary_2021} for RRLs in the LMC.} based, under the assumption of an axisymmetric disc, on an analysis of their total velocity dispersion and the age-velocity-dispersion relation. 

\begin{figure*}
	\includegraphics[width=18cm]{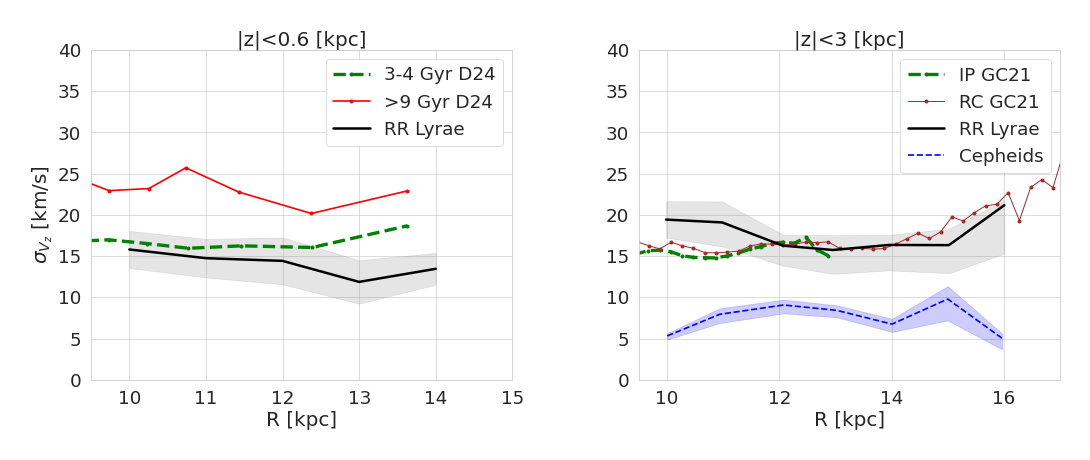}
    \caption{Both panels show the vertical velocity dispersion as a function of the galactocentric radius. Left: Our results for the RRL (black curve) in the same vertical range as the result for $3-4$~Gyr (green dashed curve) and $>9$~Gyr (red curve) by \citep[D24]{Das_sigmaVz_2024}. Right: $\sigma_{V_z}$ for the total TnD RRL sample (black curve), for the Cepheids sample (blue dashed curve), for the Red Clump (red curve) and intermediate age population (green dashed curve) from \citep[GC21]{GEDR3_anticentre_2021}}
    \label{fig:sigmas_lit}
\end{figure*}

Here, we use the vertical velocity dispersion $\sigma_{V_z}$ as a proxy of stellar age \citep[][and references therein]{Fede_spec_age_APOGEE_2023}. By using $\sigma_{V_z}$ instead of the total velocity dispersion as \cite{Iorio_RRL_thin_disc_2021}, we do not assume axisymmetry of the disc, which we have shown does not hold because of the warp. Figure~\ref{fig:sigmas_lit} shows the $\sigma_{V_z}$ as a function of $R$, in the left panel comparing, for $|Z|<0.6$~kpc, our TnD RRL sample (black curve) to the RC stars from \citet{Das_sigmaVz_2024} separated into stars with ages between $3-4$~Gyr (green curve) and $>9$~Gyr (brown curve). The velocity dispersion of our TnD RRL sub-sample is \emph{inconsistent} with the old population from \cite{Das_sigmaVz_2024}, even within the uncertainties, and, instead, shows much better agreement with the intermediate-age population. Although our RRLs show colder kinematics for $R>13$~kpc than the intermediate age, the inconsistency with the older population by \cite{Das_sigmaVz_2024} remains. We have also compared the $\sigma_{V_z}$ from our RRL sample with results from \cite{Fede_spec_age_APOGEE_2023} in the same vertical and radial range. We found them to be consistent with an age close to $3-4$~Gyr ($\sigma_{V_z}\approx15$~\kms) and also, within uncertainties, inconsistent with an age of $>10$~Gyr as the canonically old RRL population. 

The right panel of Fig.~\ref{fig:sigmas_lit} shows the $\sigma_{V_z}$ of our total sample of TnD RRL (black), our sample of Cepheids \citep[blue,][]{Cabrera_Gadea_etal_2024}, the intermediate age population (IP) and RC from \cite{GEDR3_anticentre_2021} in dashed green and red solid curves, respectively. Our TnD RRL are consistent with the IP and RC populations in $\sigma_{V_z}$ at all radii. Given that an intermediate-age population dominates the RC \citep{Red_Clump_stars_review_2016}, this adds evidence to the inconsistency of TnD RRL with a canonically old population. We remark that the RRL and the RC share the same mean vertical velocity and vertical velocity dispersion, suggesting that both have experienced the same dynamic history. 

\section{Conclusions}\label{s:conclusion}

We have shown for the first time that RRL stars trace a warped thin disc. We used a compilation of the largest catalogues of Milky Way RRL stars (Gaia SOS, ASAS-SN-II and PS1) and, using precise distances ($\sim 5\%$) and proper motions from Gaia DR3, in the anticentre direction we find a clear kinematic separation between halo ($V_\phi>-\VphiLim$) and disc ($V_\phi<-\VphiLim$) RRL. Our kinematic selection of TnD RRL stars have led us to the following conclusions:

\begin{itemize}
    \item Metal-rich RRL stars clearly exhibit the spatial and kinematic signal of the warp across the disc.  
    \item In the anticentre direction, the RRL warp in $Z$ is systematically below the Cepheid's warp by $\approx0.2$~kpc. 
    \item Also in the anticentre, the RRL warp reaches similar vertical velocities $V_z\approx 7$~\kms\ as the Cepheid's warp, but with a different radial profile. The vertical velocity of RRL as a function of the radius is similar to the results for the intermediate-age population and RC stars by \cite{GEDR3_anticentre_2021}.
    \item The pattern speed of the RRL warp is prograde with a mean value of $13\pm2$~\kmskpc\  for $R>12$~kpc, similar to previous results on the pattern speed for other stellar populations.
    \item The vertical velocity dispersion for our kinematically selected RRL is consistent with intermediate age population $3-4$~Gyr. In contrast, we find that the expected velocity dispersion of a population with the ages typically associated with RRL (>10Gyr) is significantly higher, supporting a dynamical intermediate age for our sample of TnD RRL stars.
    \item The vertical velocity of the RRL and the RC stars reported by \cite{GEDR3_anticentre_2021} shows the same trend in their mean and dispersion, reflecting their similar dynamical history.
\end{itemize}

These results open a new window with which to analyse the Galactic disc's warp. The addition of a new standard candle tracing the warp across the entire disc motivates its use to study wave phenomena in the TnD. Future works about the Milky Way warp's origin will find in the RRL population the characterisations of $Z$ and $V_z$ needed to understand how this wave differs with the stellar population age.

   
\begin{acknowledgements}
MC and CM thank Adrian Price-Whelan, Eugene Vasiliev, Payel Das, Paul J. McMillan and Friederich Anders for useful discussions and are grateful for the hospitality and support of the CCA at Flatiron Institute where part of this research was carried out.
This research has been supported by funding from project FCE\_1\_2021\_1\_167524 of the Fondo Clemente Estable, Agencia Nacional de Innovaci\'on e Investigaci\'on (ANII). This work has made use of data from the European Space Agency (ESA) mission {\it Gaia} (\url{https://www.cosmos.esa.int/gaia}), processed by the {\it Gaia} Data Processing and Analysis Consortium (DPAC, \url{https://www.cosmos.esa.int/web/gaia/dpac/consortium}). Funding for the DPAC has been provided by national institutions, in particular the institutions participating in the {\it Gaia} Multilateral Agreement.\\  

{\it Software:} Astropy \citep{astropy2018},
    AstroML \citep{astroML},
    Matplotlib \citep{mpl},
    Numpy \citep{numpy},
    Jupyter \citep{jupyter2016}, 
    TOPCAT \citep{Topcat2005,Stilts2006} and
    PYGAIA (\url{https://github.com/agabrown/PyGaia}).

\end{acknowledgements}

\appendix

\section{Rejected stars}\label{a:rrl_rejected}

After visual inspection of the light curves of the kinematically-selected sample of TnD RRL stars in the anticentre (Sec.~\ref{s:SelecAntiCentre}), there were 16 RRLs rejected as misclassified. These are summarised in Table~\ref{t:rrl_rejected}. The majority of rejected stars are eclipsing binaries, most of type EB ($\beta$~Lyrae or semi-detached) that were misclassified as \rrc\ stars in the Gaia DR3 SOS catalogue. We have indicated with P=2Psos when a true period twice of that reported by SOS produces a better light curve (with both eclipses clearly visible). In all of these cases, when available, the ratios between the G, BP and RP bands are close to unity, further supporting their classification as eclipsing binaries.

\begin{table}[]
\small
\begin{center}
\begin{tabular}{cl}
\hline\hline
Gaia DR3 \verb|source_id| & Comment \\
\hline
51251471083601920   & Eclipsing (EB), P=2Psos   \\
201773650856861696  & Eclipsing (EB), P=2Psos   \\
87159184382553088   & Eclipsing (EB), P=2Psos   \\
182118845420393088  & Eclipsing (EB), P=2Psos   \\
412435405816698880  & G-band too sparse, P=2Psos   \\
412572810414744064  &  Eclipsing (EB), P=2Psos   \\
248905041414230656  &  Eclipsing (EB), P=2Psos   \\
258626781489990272  &  Noisy, very low ampliture (0.2 in G)   \\
513743720002047488  &  Noisy, double sequence   \\
668027908865948032  &   Eclipsing (EW), P=2Psos   \\
3100435740214782080 &  Too sparse, unlikely shape   \\
2208678999172871424 &  Not an RRL. G/BP/RP well sampled\\ 
&too red, amplitude ratios inconsistent  \\
3108887445579879552 &  Eclipsing (EB), P=2Psos   \\
3112279365934238720 &  Probably Eclipsing (EA)   \\
3099365949760856448 &  Eclipsing (EB/EW), P=2Psos   \\
3324366332775173760 &  Probably Eclipsing (EA)   \\
\hline
\end{tabular}
\caption{RRL stars rejected based on their light curves. Eclipsing binary types are denoted EA (detached), EB (semi-detached) and EW (contact)}
\label{t:rrl_rejected}
\end{center}
\end{table}

\section{RRL Selection Function}\label{a:RZ_selection_function}

\begin{figure}
    \centering
    \includegraphics[width=\linewidth]{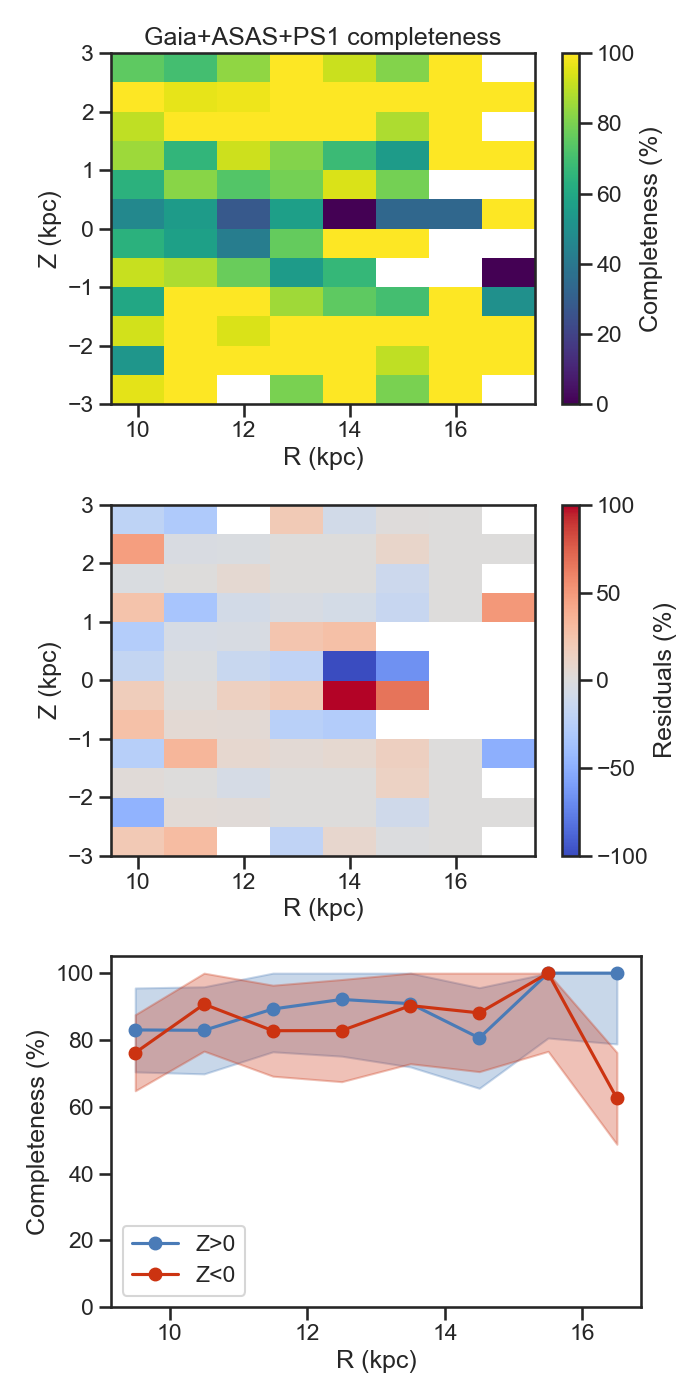}
    \caption{Gaia+ASAS+PS1 RRL catalogue completeness. Top: completeness in the $Z$ vs $R$ plane. Middle: Residuals in the $Z$ vs $R$ plane. Bottom: Completeness as a function of $R$ for $Z>0$ (blue) and $Z<0$ (red). The shaded regions correspond to $1\sigma$ uncertainties.   }\label{f:selection_function}
\end{figure}

Figure~\ref{f:selection_function} shows the selection function for the composite Gaia+ASAS+PS1 RRL catalogue as inferred by \citet{Mateu2024}, based on the method described in \citet{Mateu2020}. The top panel shows the completeness of the RRL catalogue in the $Z$ versus $R$ plane. The middle panel shows the residuals obtained by subtracting from the completeness map a north-south flipped version of itself. The bottom panel shows the completeness as a function of $R$, separately for the north (blue) and south (red) hemispheres. 

These maps illustrate there are no systematics in the selection function favouring a detection of RRL in any hemisphere versus the other. The map shows that although completeness starts to decrease towards the Galactic plane for the outermost radii ($R>16$~kpc), it remains symmetric with respect to the Galactic midplane. It is worth stressing that these estimates are empirical and not based on any Galactic model and, so, there is no prior assumption of symmetry. 

\section{The warp in the anticentre}\label{apend_AC}

\begin{figure}
	\includegraphics[width=9cm]{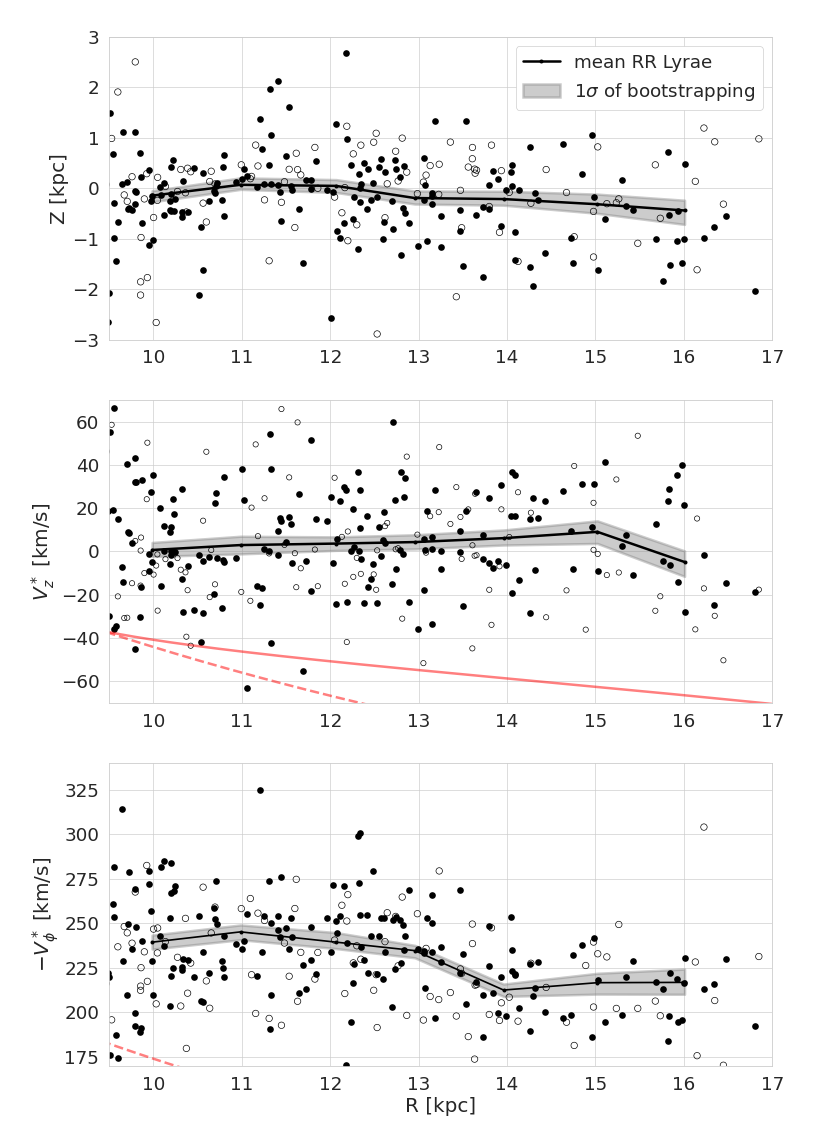}
    \caption{The three panels show the mean as a function of the radius, the vertical height $Z$ (top panel), the vertical velocity $V_z^*$ (middle) and the azimuthal velocity $V_{\phi}^*$ (bottom). In all panels the black dots are the RRL stars with photometric metallicity and the empty dots are those with metallicity assigned. The black line is the mean computed in 1~kpc radial bins in each panel. The grey shaded area corresponds to $1\sigma$ of the $300$ bootstrapping realisations of the mean. The red curve shows the track of Sagittarius \citep{Pau_sag_2022} in the anticentre above the plane (solid curve) and below the plane (dashed line).}
    \label{fig:Vz_R_disc}
\end{figure}

Figure~\ref{fig:Vz_R_disc} shows $Z$ (top), $V_z^*$ (middle) and $V_{\phi}^*$ (bottom) as a function of the radius, with the black lines corresponding to the same means and uncertainties shown in Fig.~\ref{fig:Vz_Z_lit}. The black dots are the RRL stars with photometric metallicity and the empty dots are those with metallicity assigned. Because RRLs trace substructures such as the Sagittarius \citep[Sgr,][]{Pau_sag_2022} and Monoceros \citep[][]{Pau_Mon_2021} streams, we checked whether these were contaminating our disc sample in the anticentre direction. The red curve shows Sagittarius track above the Galactic plane (solid curve) and below the plane (dashed curve). \
Our cut in $V_\phi^*$ to select disc RRL removes most of Sagittarius for $R>10$~kpc and the cuts in $|Z|<3$~kpc and $9.5<R/\text{kpc}<17$ guarantee its absence in our final disc sample. Also we checked that none of the Sagittarius RRL from \citet{Pau_sag_2022} match with our disc sample. For Monoceros, we do not find any RRL from our disc sample to trace the track reported by \cite{Pau_Mon_2021}. 

\section{Pattern speed of the warp}\label{a:PatternsSpeed}

The mean vertical velocity of the stars $V_z$ and the mean vertical height $Z$ are related by \citep[see eq. 20][]{Chequers_Widrow_2018_Spontaneous}

\begin{equation}
V_z(R,\phi,t)=\frac{\partial Z}{\partial t}+\frac{\partial Z}{\partial \phi}\Omega \label{eq:Z_dot} 
\end{equation}

where we have ignored the contribution of the radial velocity by the same argument given in \citep{Cabrera_Gadea_etal_2024}. The partial derivative with respect to time in Eq.~\ref{eq:Z_dot} contains all the contribution to $V_z$ only by the changes in the warp structure given by its time evolution (patterns speed and changes in amplitudes of all modes).  
We will assume that the warp only rotates and all of its modes have the same pattern speed $\omega$, then $\partial_t Z=-\omega\partial_\phi Z$. Given the similarity in $Z$ and $V_z$ in phase with that of the Cepheids shown in Fig.~\ref{fig:XY_Z} seems reasonable to assume that $\dot{A}=0$. We therefore can calculate $\omega$ as

\begin{equation}
    \omega=\Omega- \frac{V_z}{{\partial_\phi Z}}
    \label{eq:omega}
\end{equation}

In order to evaluate $\partial_\phi Z$ we fit in each radial bin a $m=1$ mode and calculate its derivative. We measure $\Omega$ as the mean $V_\phi^*$ divided by the radius of the radial bin. Because we can apply Eq.~\ref{eq:omega} in each radial bin, we get the pattern speed as a function of the radius.

%
%

\bibliographystyle{aa}
\bibliography{warp_refs}

\end{document}